# Political, economic, and governance attitudes of blockchain users

Lucia M. Korpas[1], Seth Frey[2,3], Joshua Tan[1,4*]

[1]The Metagovernance Project, Brookline, MA, USA

[2]Department of Communication, University of California Davis, Davis, CA, USA

[3]The Ostrom Workshop, Indiana University, Bloomington, IN, USA

[4]University of Oxford, Oxford, Oxfordshire, UK

**\* Correspondence:**
Joshua Tan
joshua.z.tan@gmail.com

**Abstract**

We present a survey to evaluate crypto-political, crypto-economic, and crypto-governance sentiment in people who are part of a blockchain ecosystem. Based on 3710 survey responses, we describe their beliefs, attitudes, and modes of participation in crypto and investigate how self-reported political affiliation and blockchain ecosystem affiliation are associated with these. We observed polarization in questions on perceptions of the distribution of economic power, personal attitudes towards crypto, normative beliefs about the distribution of power in governance, and external regulation of blockchain technologies. Differences in political self-identification correlated with opinions on economic fairness, gender equity, decision-making power and how to obtain favorable regulation, while blockchain affiliation correlated with opinions on governance and regulation of crypto and respondents' semantic conception of crypto and personal goals for their involvement. We also find that a theory-driven constructed political axis is supported by the data and investigate the possibility of other groupings of respondents or beliefs arising from the data.

## 1  Introduction

As blockchain technology has evolved over more than a decade, cryptocurrencies and crypto-economic systems have had a growing impact on the world. Millions of people have involved themselves in crypto[1]: as of 2021, around 15 percent of American adults have reported owning cryptocurrency (Perrin 2021), and many other countries have even higher adoption rates (Buchholz 2021). The past few years have seen the growth of decentralized apps and the crypto startup industry. Correspondingly, governments are beginning to take regulatory actions. Also, even as blockchain ecosystems move towards less computationally-intensive consensus mechanisms, the ongoing environmental impact of blockchain use is huge.

Given the impact of crypto-economic activity on individuals and on shared resources, it is increasingly important to understand how its users are relating to the technology. While the hard data of cryptocurrency transactions and account balances is often publicly available by design, users' motivations for engaging with crypto are more opaque. There is little existing data on the stated beliefs or attitudes of the variety of people using blockchain technologies. What do blockchain users believe about the economic, political, and social relevance of crypto? While there has been attention to the attitudes of the general population towards cryptocurrencies and blockchain technology (Perrin 2021; "Global State of Crypto, 2022" 2022), there is also a need to understand the beliefs of active participants of blockchain ecosystems.

---

[1] Throughout the text, we use the term "crypto" to encompass blockchain technologies such as cryptocurrencies and the communities and ideologies which drive their development and use.

What do blockchain ecosystem participants believe about how the technology is being – or should be – developed, used, and regulated? Are there discrete types of crypto contributors, or is there a spectrum of beliefs? What specific beliefs are most relevant in distinguishing respondents between types or along axes? This work is a first step in the development of a framework for thinking about this spectrum or grouping of beliefs in crypto.

We report the results of a large-scale survey of participants in the blockchain economy. The survey was designed to shed light on respondents' socioeconomic and sociopolitical beliefs relating to crypto, economic modes of engagement with crypto, and attitudes towards governance of blockchain technology. We describe the distributions of these responses and their relationships to self-reported political ideology and specific crypto ecosystems such as Bitcoin and Ethereum.

We also evaluate the survey instrument itself: are the questions able to assess distinct and relevant facets of beliefs? Can we identify underlying factors which describe broader groupings of beliefs? Using factor analysis methods, we find that a political axis and corresponding typology, informed by the Pew Research Center's Political Typology Quiz, meaningfully describes variation between respondents.

## 2    Background

While there is no existing political theory of crypto per se, there are substantial ethnographic studies of crypto communities (and related digital communities) that address the political dimensions of crypto. For example, ethnographic studies have informed the creation of a proposed political typology of blockchain projects (Husain, Franklin, and Roep 2020), reflecting earlier ideas on the "intrinsic" political values of technical artifacts (Winner 1980). In this vein, cryptocurrencies have been characterized as realizations of crypto-anarchist values such as privacy and autonomy (Chohan 2017; Beltramini 2021), following in the footsteps of earlier cypherpunk writings (Hughes 1993; May 1994) as well as the original Bitcoin whitepaper (Nakamoto 2008). Other ethnographies have described issues of on- and off-chain governance (De Filippi and Loveluck 2016) and the political motivations and cultural context of projects such as Bitcoin (Golumbia 2016) and Ethereum (Brody and Couture 2021).

A previous industry survey, conducted by CoinDesk in 2018, contained several questions related to politics and governance (Ryan 2018; Bauerle and Ryan 2018), though the questions focused more specifically on individual projects and topical questions such as reactions to SEC rulings on the securitization status of Ethereum.

Distinct from questions about political values, the topic of blockchain governance—including the relationship between blockchains and traditional governments—is one of the most salient and polarizing questions in crypto, one that has led to the creation, forking, and dissolution of many projects. While we cannot recount all the major positions here (some of which are reflected in the survey itself; see "Methodology"), there is a broad distinction between approaches that emphasize on-chain governance and those that emphasize off-chain governance. A number of academic analyses have studied these different approaches to blockchain governance (Reijers, O'Brolcháin, and Haynes 2016; Liu et al. 2021; van Pelt et al. 2021), along with a vastly greater number of industry manifestos and opinion pieces (Zamfir 2019; Szabo 1996).

## 3 Methodology

### 3.1 Survey questions

The survey consists of 19 questions related to respondents' crypto-related beliefs and activities, with three types of questions interspersed: those eliciting opinions about the political dimensions of crypto activity ("crypto-political"), those eliciting economics opinions ("crypto-economic"), and those eliciting attitudes about the governance of crypto projects ("crypto-governance"). All questions were multiple choice, with 2-4 possible selections, and the respondent could opt not to answer. See Table 1 for the full list of questions.

The survey questions and provided choices included both a formal portion drawing from existing political survey instruments and a more exploratory portion intended to elicit beliefs relevant to a general crypto-political typology. In particular, a few of the questions selected (Q11-13, Q15, Q19), were based on questions from Pew's Political Typology Quiz (Nadeem 2021) and intended to relate to political sentiment. Other questions (e.g., Q1, Q17) were developed in collaboration with a number of community members in crypto, drawing on the culture, memes, and references common in crypto. Altogether, the content was designed to elicit respondents' primary modes of economic engagement with crypto, their political sentiment, and opinions as to how crypto communities themselves should be governed.

### 3.2 Construction of political "types" and identification of "axes" of belief

Our choice to identify separate "axes" of economic, political, and governance beliefs were based on discussion with community members and in analogy to existing classifications such as the traditional "left-right" political axis. For one of these, the political axis, we also leveraged our study design to group and relate questions more directly by defining a continuous construct intended to assess respondents' crypto-political leanings. We identified a subset of questions as most relevant to political orientation, and computed a score for each participant by summing the responses to these questions (coded with values in the range [-1, 2] as described in Table 1) in analogy to the Pew methodology (Nadeem 2021). The lowest and highest scores on this political "axis" were designed to highlight extreme positions of collectivist and anarcho-capitalist approaches to using blockchain technology. Five discrete types were defined by thresholds in the score according to Table 2: crypto-anarcho-capitalist, crypto-libertarian, centrist, crypto-communitarian, and crypto-leftist; these types were developed both with definitions from the Pew typologies and with input from the community.

### 3.3 Recruitment

We relied on a convenience (self-selected) sample of participants in the crypto community. Participants were recruited by distributing the survey through blockchain-focused forums and listservs, conferences (LisCon and ETHDenver), social media posts, and articles published on blockchain-focused news sites.

We motivated voluntary participant engagement with two strategies. We presented the survey as a quiz that assigned respondents one out of an entertaining typology of "types" on the basis of their responses ("crypto-leftist," "cryptopunk," etc.) immediately upon completion of the survey. Stylized as "factions", the crypto-political types corresponded to the political types we defined based on the Pew typology, while the crypto-economic and crypto-governance types were constructed by using thresholds to partition respondents into five ad hoc types (for more detail, see Section 1.1 of the Supplementary Material). We also incentivized survey completion with the opportunity to receive a non-fungible token (NFT) corresponding to their assigned "type", contingent upon their provision of a valid Ethereum wallet address or ENS name.

### 3.4 Analysis

To survey the overall landscape of crypto-political beliefs, we observed the distribution of choices selected by respondents. We aggregated these responses for each question, including the null response of no choice selected, and computed the margins of error for a 95% confidence interval, assuming a random sample of the population.

To investigate how political self-identification and participation in specific blockchain ecosystems related to beliefs, we grouped participants by their responses to the corresponding questions. We then determined which questions displayed a statistically significant difference in the distribution of responses between these groups.

We also wanted to understand which questions were most meaningful in differentiating respondents. To this end, we first performed a check on the extent to which each question measured a distinct belief by computing the correlation between responses to different questions as Cramer's V (a version of Pearson's chi-squared statistic scaled to provide a measure of association). Then we used principal component analysis (PCA) to identify which of the 48 choices provided across 17 questions explained the most variance between respondents. Specifically, we looked at the component loading for each feature, i.e., contribution to the first principal component.

For use with PCA, we normalized the feature data (shifted to a mean of 0 with unit variance). Note that for questions where only two choices were provided, the alternative answer contributed with equal magnitude (though opposite sign). We omitted questions 2 and 19, relating to specific blockchain ecosystems. We also omitted answers from respondents that did not answer all questions.

We also wished to evaluate to what extent our crypto-political types, delineated from the political score we defined, corresponded with patterns in beliefs across respondents. From the PCA results, we can identify to what extent the assigned types or classes directly correlate with any of the first few components.

## 4 Results

### 4.1 Responses and respondents

Between September 27, 2021 and March 4, 2022, the survey received 3710 responses. In 3418 (92%) of these, all questions were answered. For questions presented to all respondents, the percentage of respondents who chose not to answer each question was between 0.5% and 1.5% across all questions. The survey took on average 8 minutes and 40 seconds to complete.

### 4.2 Responses to questions: political, economic, and governance attitudes

Overall, respondents were varied in their perceptions of the distribution of economic power in crypto and their personal attitudes towards crypto. They were also split between the most common responses to two questions on the distribution of power in governance of crypto. There was somewhat more agreement on broad beliefs towards external regulation of crypto, though respondents disagreed on some of the specifics and in matters of degree. The largest majorities were observed in questions relating to the social implications of crypto.

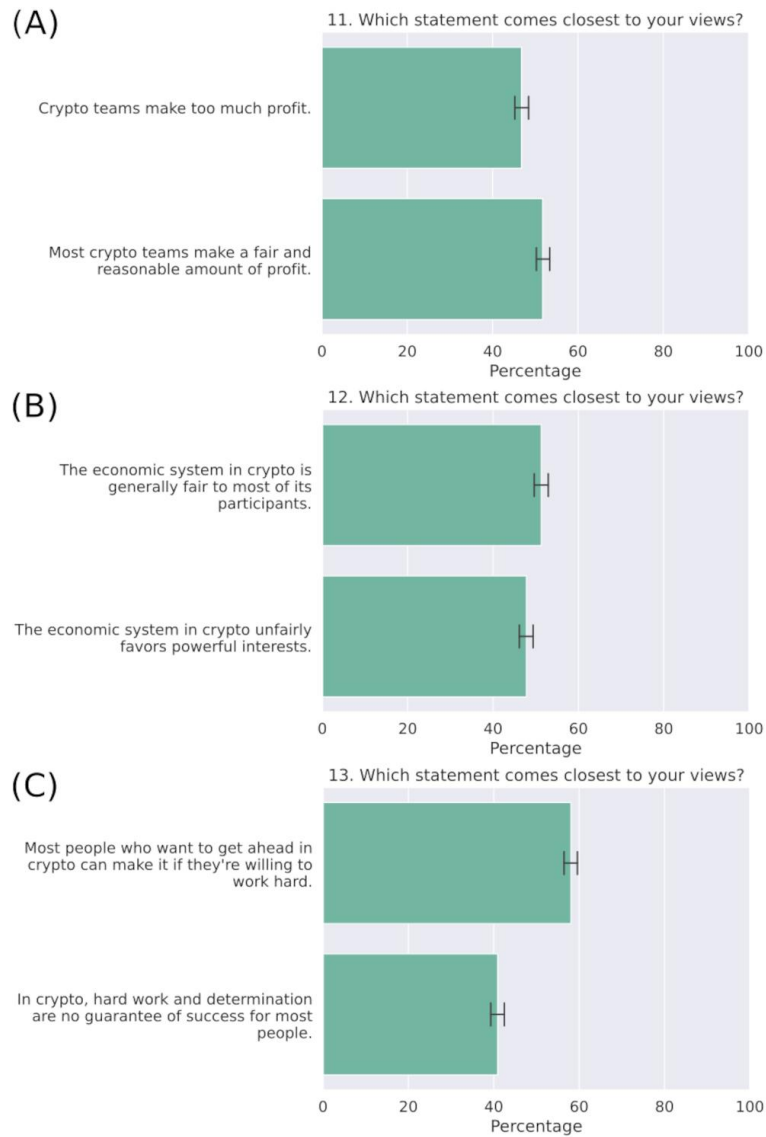

**Figure 1.** Responses to **(A)** question 11, **(B)** question 12, and **(C)** question 13 on perceptions of the distribution of economic power in crypto, with 95% confidence intervals. Though optimistic beliefs about the current state of crypto-economics were slightly more prevalent, dissatisfaction with the fair distribution and attainability of crypto-economic wealth was nearly as frequent.

Perceptions of the distribution of economic power in crypto were closely split between the two choices provided for each question (Figure 1). By a few percentage points, a slightly higher proportion of respondents believed that most crypto teams make "a fair and reasonable amount of profit" rather than "too much profit" (Q11, Fig. 1(A)) and that the economic system in crypto "is generally fair to most of its participants" rather than "unfairly favors powerful interests" (Q12, Fig. 1(B)). A majority (58%) believed that "most people who want to get ahead in crypto can make it if they're willing to work hard" (Q13, Fig. 1(C)).

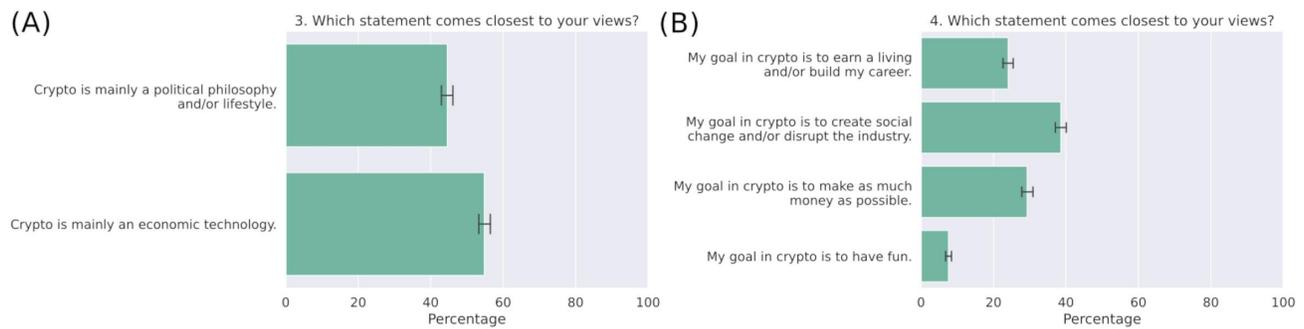

**Figure 2.** Responses to **(A)** question 3 and **(B)** question 4 on personal attitudes towards blockchain, with 95% confidence intervals. Together, these responses show that both a desire for sociopolitical change and an interest in personal financial gain were common factors in participants' interest in blockchain technologies.

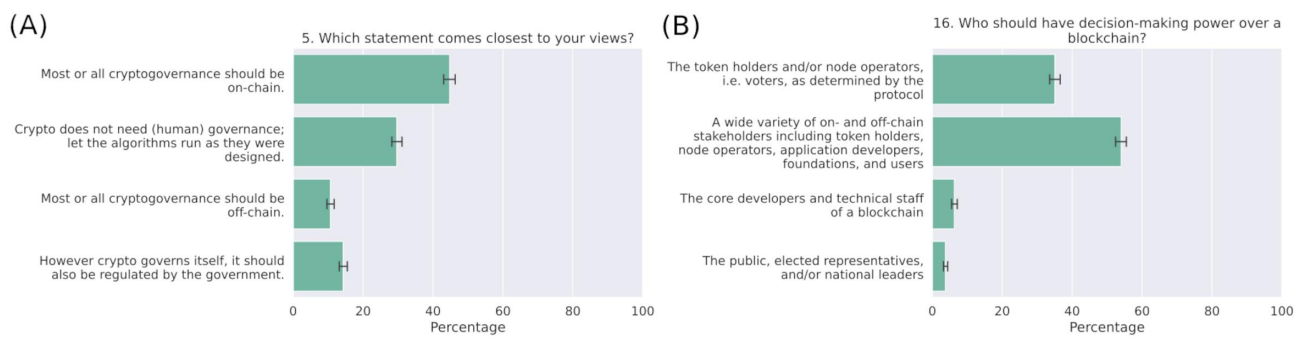

**Figure 3.** Responses to **(A)** question 5 and **(B)** question 16 on blockchain governance, with 95% confidence intervals.

Personal attitudes towards crypto were also diverse (Figure 2). Respondents were divided on whether they regarded crypto as "mainly a political philosophy and/or lifestyle" or "mainly an economic technology", with a slight majority favoring the latter (Q3, Fig. 2(A)). There was no majority in respondents' goals for their own involvement in crypto: the most common goal was "to create social change and/or disrupt the industry" (39%), followed by "to make as much money as possible" (29%) (Q4, Fig. 2(B)).

Normative beliefs about the distribution of power in governance of crypto appear to be in some tension (Figure 3). Most respondents favored a hands-off approach to the governance of crypto, with 45% believing that "most or all cryptogovernance should be on-chain" and 30% believing "crypto does not need (human) governance" (Q5, Fig. 3(A)). However, a majority of respondents believed that "a wide variety of on- and off-chain stakeholders" should have decision-making power over a blockchain (though the next most common response was "the token holders and/or node operators, i.e., voters, as determined by the protocol") (Q16, Fig. 3(B)). Note that while the most common responses to each of these questions are not incompatible, their coexistence indicates a possible tension in the community between maximizing on-chain governance and empowering off-chain stakeholders.

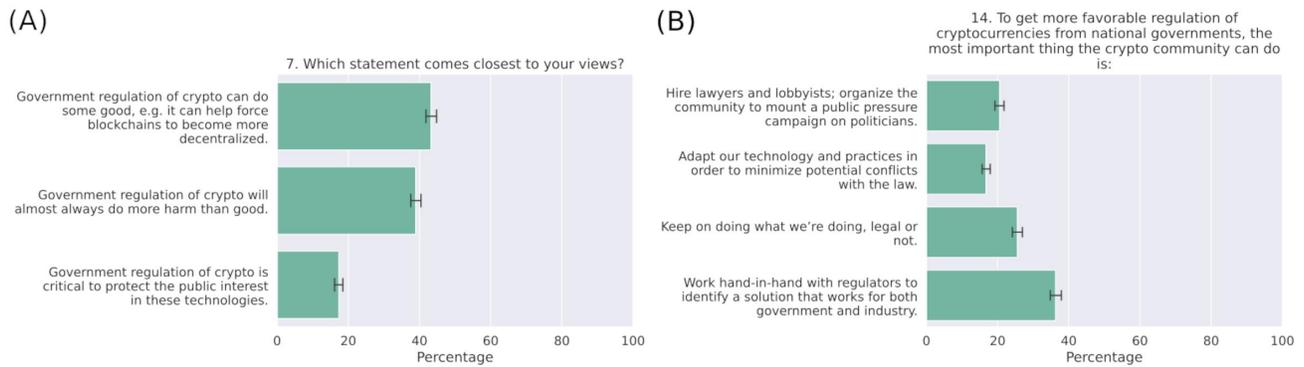

**Figure 4.** Responses to **(A)** question 7 and **(B)** question 14 on external regulation of blockchain technologies, with 95% confidence intervals.

Regarding external regulation of blockchain technologies, respondents were somewhat more consistent (Figure 4). A majority of respondents believed at least some good will come of government regulation of crypto, though nearly 40% asserted that "government regulation of crypto will almost always do more harm than good" (Q7, Fig. 4(A)). In line with the above, when asked what the most important thing the crypto community can do to get more favorable regulation of cryptocurrencies from national governments, a plurality of respondents sought a cooperative relationship with government, choosing to "work hand-in-hand with regulators to identify a solution that works for both government and industry," versus adopting an evasive approach to "adapt our technology and practices in order to minimize potential conflicts with the law" or even an antagonistic one to "mount a public pressure campaign on politicians" or to "keep doing what we're doing, legal or not" (Q14, Fig. 4(B)). Also, more than three-quarters of respondents believed that "having a central bank run a cryptocurrency is a bad idea" (Q8). Overall, though a majority of respondents were willing to accept or even collaborate on regulation, large minorities strongly disagreed, and distaste for direct government involvement in implementations of crypto technology was common.

On the social implications of crypto, most respondents were in agreement, believing that blockchain and DeFi are "beneficial technologies that, on balance, will help most members of society" (Q10). Even so, more than a quarter of respondents believed that crypto "has a gender problem" (Q15). Also, around a quarter of respondents indicated privacy is "the most important feature of blockchain and crypto" (Q6).

We asked two additional questions on political orientation and blockchain ecosystem affiliation (Figure 5). Only 14 percent of respondents considered themselves "conservative or right-wing" (532 respondents) with the remaining participants split equally (with no statistically significant difference) between "liberal or left-wing" (1550) and "neither" (1599; Q18, Fig. 5(A)). Nearly all participants (97%) stated an affiliation with at least one blockchain ecosystem or community (Q19, Fig. 5(B)), supporting our use of this dataset to focus on users of blockchain technology (rather than the general public). In particular, of the 3591 respondents who indicated affiliation with at least one blockchain, 2175 (61%) selected affiliation with Ethereum and 1120 (31%) with Bitcoin (Fig. 5b). Note that these are not mutually exclusive groups (789 indicated affiliation with both); furthermore, though a majority of respondents only specified one affiliation, less than a quarter believe that "there is one (layer 1) blockchain that is the best" (Q1). In the following two subsections, we discuss the relation of these distributions with respondents' beliefs in more depth.

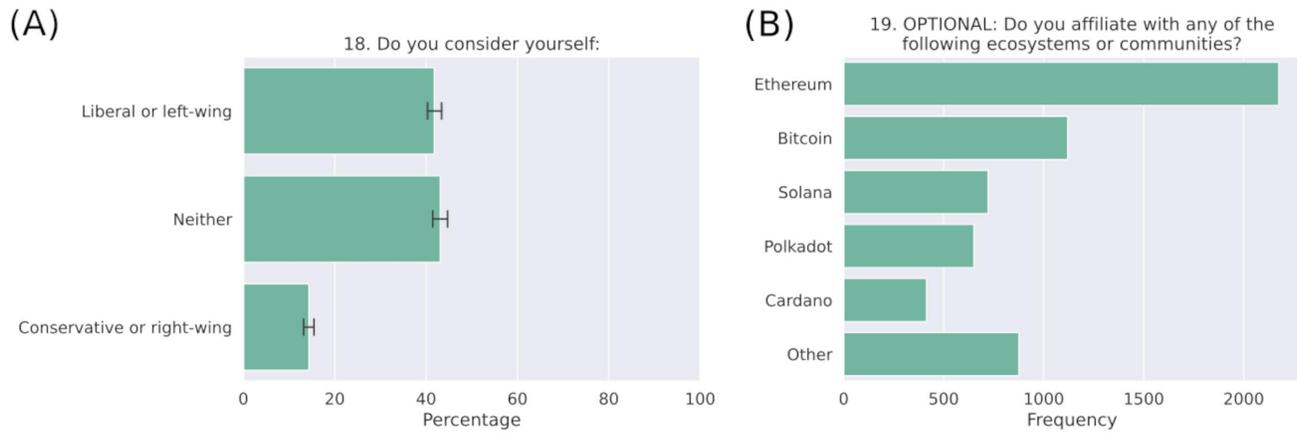

**Figure 5.** Responses to **(A)** question 18 on political orientation and **(B)** question 19 on blockchain ecosystem affiliations, with 95% confidence intervals.

The distribution of responses for the questions not covered in this section are included in the Supplementary Material (Supplementary Figures S1-S4).

**4.3  Differences between respondents by self-reported political orientation**

To examine the differences in opinion between the left-of-center, right-of-center, and unaligned groups, we compared the distribution of answers selected by respondents affiliated with each group (Q18). We found that perceptions of economic fairness and gender equity elicited the clearest differences between the three political orientation groups, with economic fairness especially differentiating left-of-center respondents from the other two groups. Beliefs about governance, regulation, and personal goals in crypto differentiated right-of-center respondents from the other two groups. Differences between political orientation groups were ubiquitous: all but one question had at least one statistically significant difference between the responses groups.

The economic fairness questions (Q11, Q12, and Q13) were among those with the greatest differentiation between the three groups. Somewhat surprisingly, unlike nonaligned and right-of-center respondents, a majority of left-of-center respondents believe that "most crypto teams make a fair and reasonable amount of profit" (Q11) and "the economic system in crypto is generally fair to most of its participants" (Q12, Fig. 6). Though a majority of both right-of-center and nonaligned respondents believed instead that "the economic system in crypto unfairly favors powerful interests", right-of-center respondents were more likely than nonaligned respondents to choose this answer (Q12). However, left-of-center respondents were more likely than right-of-center or nonaligned respondents to believe that "hard work and determination are no guarantees of success" in crypto (Q13).

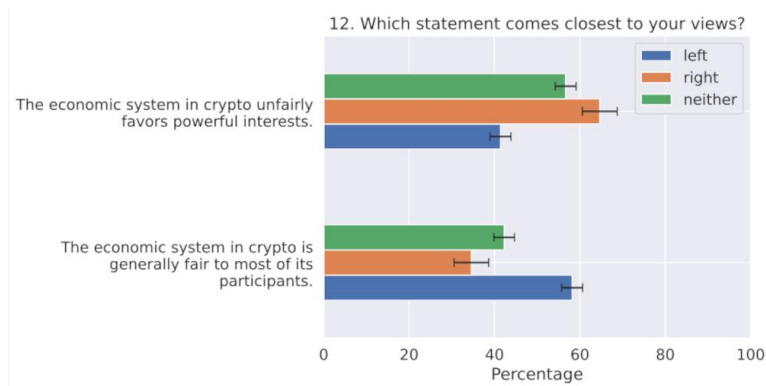

**Figure 6.** Responses, grouped by self-reported political affiliation, to question 12 on crypto-economic fairness, with 95% confidence intervals. Taken together with questions 11 and 13, this distribution shows that left-of-center respondents overall held a different set of beliefs about wealth distribution and economic opportunity than other respondents.

Question 12 was one of three questions for which all three groups had a statistically different distribution of responses. Another was on gender equity: right-of-center respondents were least likely to believe "crypto has a gender problem," nonaligned respondents somewhat more likely, and left-of-center respondents most likely, with about half of left-of-center respondents selecting this answer (Q15). This spread shows that self-reported political alignment relates to not only economic but also social issues in the use of blockchain technology.

Differences also arose between the groups in the most common answer to questions on decision-making power and how to obtain favorable regulation. When asked who should hold decision-making power over a blockchain, right-of-center respondents were more likely to choose "the token holders and/or node operators" than "a wide variety of on- and off-chain stakeholders"; the reverse was true for left-of-center and nonaligned respondents, with left-of-center respondents more likely than other respondents to choose a variety of stakeholders (Q16, Fig. 7). Concerning how to obtain favorable regulation, left-of-center and nonaligned respondents were most likely to choose "work hand-in-hand with regulators" out of the available choices, and more likely to do so than right-of-center respondents; in contrast, right-of-center respondents were, within confidence intervals, evenly split between three of the four available choices (Q14).

Other statistically significant differences occurred in the distribution of responses, where one of the three groups differed from the other two. Right-of-center respondents were most likely to choose "make as much money as possible" as their goal and less likely to select "create social change and/or disrupt the industry"; the reverse was true for left-of-center and nonaligned respondents (Q4). However, left-of-center respondents were less likely than others to believe crypto needs to prioritize "building art and community" to grow (Q9). Also, a smaller proportion of left-of-center respondents than other respondents believed that privacy is "the most important feature of blockchain" (Q6).

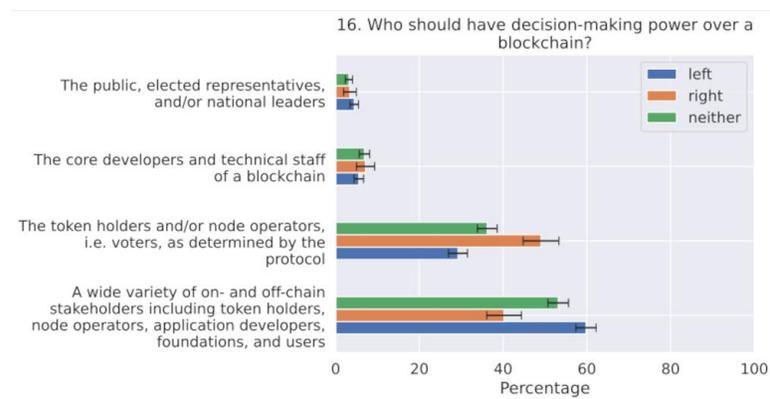

**Figure 7.** Responses, grouped by self-reported political affiliation, to question 16 on decision-making power, with 95% confidence intervals. Together with question 14, this distribution indicates that right-of-center respondents were more likely than other respondents to hold beliefs aligned with minimizing external influence on blockchain governance and development.

Left-of-center respondents were less likely to believe that "crypto does not need (human) governance," while nonaligned respondents were less likely to believe "however crypto governs itself, it should also be regulated by the government" (Q5). Left-of-center respondents were also more polarized on government regulation: they were less likely to believe it "can do some good," and more likely to believe it is either "critical to protect the public interest" or "will always do more harm than good" (Q7).

### 4.4 Differences between respondents by Bitcoin and Ethereum affiliation

At present, dynamics in the crypto community are largely driven by actors in two ecosystems: Bitcoin and Ethereum. To examine differences in opinion between the 61% of respondents affiliated with Ethereum and the 31% (non-exclusive) affiliated with Bitcoin, we compared the distribution of answers selected by respondents affiliated with each of the two blockchains. We found an overall quite similar distribution of responses regardless of affiliation, with a few statistically significant differences arising in beliefs about cryptogovernance, the semantics of the term crypto, personal goals in crypto, and stated political orientation.

Governance and regulation of crypto were a key topic distinguishing Bitcoin affiliates from Ethereum affiliates (Figure 8). Bitcoin affiliation was associated with a higher likelihood of believing that "crypto does not need (human) governance" (Q5, Fig. 8(A)) and that "token holders and/or node operators" should have decision-making power over a blockchain, whereas Ethereum was associated with "a wide variety of on- and off-chain stakeholders" (Q16, Fig. 8(B)). Somewhat surprisingly, Bitcoin affiliation was also associated with a higher likelihood of believing that government regulation of crypto "can do some good" (Q7), although there was no statistically significant difference in opinions on how to obtain favorable regulation (Q14). Thus, it appears that Bitcoin affiliation is associated with a higher rate of wanting to maximize on-chain governance but also of tolerance of external regulation, perhaps in particular that which "can help force blockchains to become more decentralized," as is included in the wording of question 7.

Respondents' semantic conception of crypto and their personal goals for their involvement also had some relation to blockchain affiliation: Bitcoin affiliation was associated with a higher likelihood of believing "crypto is mainly an economic technology" (Q3) and identifying with the statement "my goal in crypto is to make as much money as possible" (Q4). Ethereum affiliation was associated with a higher likelihood of believing that "the economic system in crypto unfairly favors powerful interests" (Q12) and that "crypto has a gender problem" (Q15).

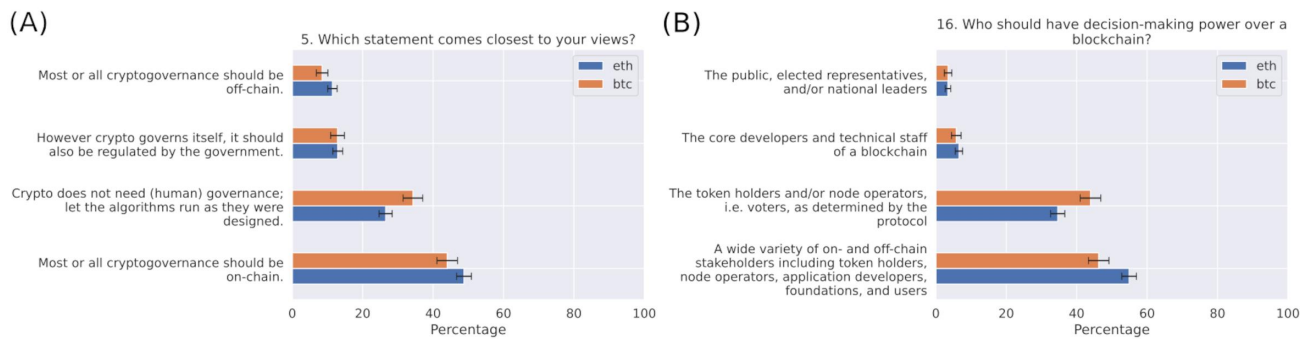

**Figure 8.** Responses, grouped by blockchain affiliation, to **(A)** question 5 and **(B)** question 16 on blockchain governance, with 95% confidence intervals. These distributions indicate that Bitcoin affiliates were more likely to favor a narrow definition of governance and its participants.

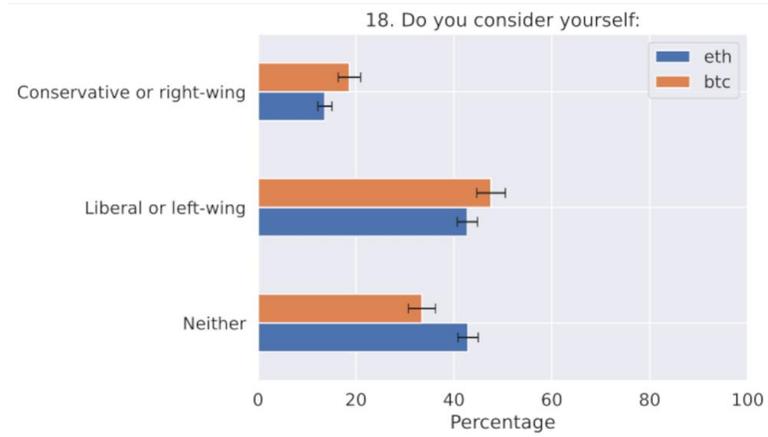

**Figure 9.** Responses, grouped by blockchain affiliation, to question 18 on self-reported political affiliation, with 95% confidence intervals. While there was a statistically significant difference between affiliates of the two blockchains in identifying as right-of-center or nonaligned, Cramer's V indicates that the strength of association between blockchain affiliation and political orientation was low.

For question 18 on political orientation, Bitcoin affiliation correlated with a higher likelihood of selecting "conservative or right-wing" and lower likelihood of selecting "neither" (Figure 9). There was no statistically significant difference between the proportions of respondents who chose "liberal or left-wing". Given that we were interested in analyzing blockchain affiliation separately from stated political orientation, we additionally checked for the strength of association between Q18 and a reduced version of Q19 with the options "Bitcoin", "Ethereum", and "Neither" (not mutually exclusive). Cramer's V was low (less than 0.15) for all combinations of responses, indicating at most very weak association between the two questions (Supplementary Figure S5). This gives us confidence that Bitcoin and Ethereum affiliation were not strongly associated with stated political orientation.

### 4.5 Validation of survey instrument

To assess any correlations between responses to different questions, we computed the correlation matrix for all pairs of questions (Supplementary Fig. S6). Of the 153 unique pairs, most showed little if any association (V < 0.1); the strength of association was weak for 52 questions (0.1 <= V < 0.3), and one question pair related to wealth distribution (Q11-Q12) showed a moderate strength of association (0.33). The prevalence of weak or no association between distinct questions supports our assertion that each question addresses a distinct facet of a respondent's beliefs or actions. This allows us to assess the relative importance of the specific statements provided in the answer choices to explain differences between respondents.

### 4.6 Feature selection and factor analysis

To identify the beliefs which most contributed to explaining variance between respondents and to test our hypothesis, we computed the PCA vectors for individual choices (features) and examined the first principal component. Beliefs above a threshold of magnitude 0.18, corresponding to the loading each response would have if all questions contributed equally to the component, were labeled as important. The features with the largest contributions to the first principal component were the following (listed in descending order of importance):

- "The economic system in crypto unfairly favors powerful interests." (Q12)
- "Crypto has a gender problem." (Q15)
- "Government regulation of crypto will almost always do more harm than good." (Q7)
- "[I consider myself] liberal or left-wing." (Q18)
- "Crypto teams make too much profit." (Q11)
- "In crypto, hard work and determination are no guarantee of success for most people." (Q13)
- "However crypto governs itself, it should also be regulated by the government." (Q5)
- "Blockchain and DeFi are predatory technologies that, on balance, will harm most members of society." (Q10)

All three questions relating to wealth distribution and economic fairness (Q11-13) contributed more to explaining variance than most other questions. Polarized opinions on government regulation (Q5 and Q7) and one specific political affiliation (Q18) also featured here. Altogether, 5 of the 8 questions that we had coded as defining an axis of political belief had a large contribution to this leading component.

The same analysis can be done for the remaining principal components. The features with the largest component loading for the next two principal components are "Privacy is the most important feature of blockchain and crypto" (Q6) for the second principal component and "Crypto is mainly a political philosophy and/or lifestyle" (Q3) for the third. These choices, and their corresponding questions, are therefore among the more salient in explaining variance between respondents.

Altogether, however, the variance explained by only the first few components was relatively low (21% for the first three components) and less than 10% was explained by the first component alone. Taken together with weak associations between questions as described in Section 4.5, this implies that the number of latent variables required to describe respondents' beliefs is large. Indeed, factor analysis using PCA and feature agglomeration yielded a null result, meaning that features did not cluster into a few interpretable groupings (see Section 1.3 of the Supplementary Material). Even so, we find that the first principal component axis corroborates a theory-first constructed axis, as described in the following section.

### 4.7 Validation of constructed crypto-political axis

For each respondent, a political score was calculated using the values in Table 1 and a type was assigned according to the score thresholds described in Table 2. The feature selection and factor analysis results can be used to evaluate the validity of this constructed crypto-political axis.

The distribution of scores and types assigned to participants who completed the survey is shown in Figure 10. On the left-of-center side, 20% of respondents were identified as "crypto-communitarian" or "crypto-leftist", while 9% of respondents were given the "crypto-centrist" label. The most commonly assigned type was the "crypto-libertarian" types, with nearly half of respondents receiving this designation; overall, right-of-center types ("crypto-libertarian" and "crypto-anarcho-capitalist") dominated with 71% of respondents. This distribution is unimodal, low skewness, and centered around the median possible score. However, because the range of possible scores was not centered around zero, we find that a majority of respondents were labeled as crypto-politically "right-of-center". For a summary of how this distribution differed with political self-identification and blockchain affiliation, Section 1.2 of the Supplementary Material.

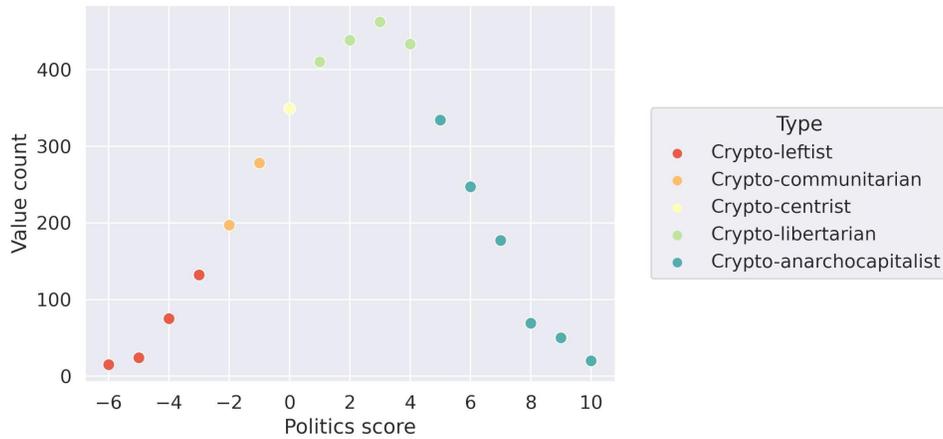

**Figure 10.** Distribution of assigned crypto-political scores and corresponding sentiment types. Despite a 14% minority of respondents identifying as ideologically conservative or right-wing, our measure placed 71% in the right-of-center libertarian and anarchocapitalist categories.

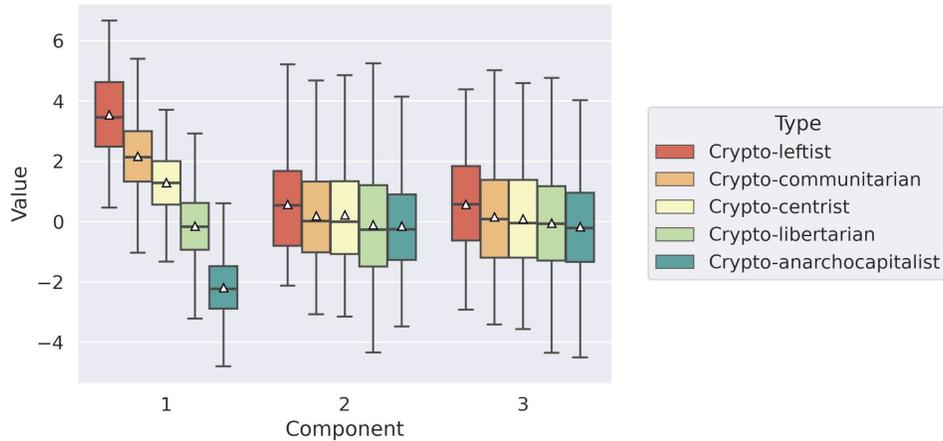

**Figure 11.** Box plots showing the distribution of values for each political type along each of the first three principal components produced by PCA, with the mean scores indicated by a white triangle. There is essentially no overlap between crypto-leftist and crypto-ancap types for component 0.

Partitioning the respondents by the types we identified for them, and plotting them in the first three PCA components, we find, again, that the first principal component succeeds at capturing the political dimension of respondent variation, while the next two components are less informative (Figure 11). The low overlap between the interquartile ranges for adjacent types indicates that the continuous construct we defined and the defined types which discretize it help to explain differences between respondents' beliefs. Thus, the constructed political axis seems to reflect true variation in the population and may be of use in future work characterizing the ideological structure of the crypto community.

## 5    Discussion

Though optimistic beliefs about the current state of cryptoeconomics were slightly more prevalent, the survey responses indicate nearly as much dissatisfaction with the fair distribution and attainability of cryptoeconomic wealth. Both a desire for sociopolitical change and an interest in personal financial gain were common factors in participants' interest in blockchain technologies. Respondents generally were optimistic about the social potential of blockchain technology, with some having reservations about its gender equity and some focusing on its privacy implications. Overall, though a majority of respondents were willing to accept or even collaborate on regulation, large minorities strongly disagreed, and distaste for direct government involvement in implementations of crypto technology was common.

Despite low rates of respondents' self-identification with "conservative or right-wing" politics, we observed a prevalence of right-of-center crypto-political types. A broadly similar distribution was observed in a CoinDesk report published in 2018 (Ryan 2018). The discrepancy between general political self-identification and our crypto-specific labeling bears further investigation. It may relate to an association of the term "conservative" with social conservatism, whereas crypto-libertarianism, the crypto-political type we found to be most common, emphasizes a form of economic libertarianism. Furthermore, the connotations of "conservative" and "liberal" vary significantly by geographic region, so the question may have been interpreted differently across respondents based on their country of residence.

The correlation of the constructed political axis with the first principal component–a commonly-used, well-validated axis–suggests a primacy of political variation in explaining patterns of responses. Furthermore, the existence of differences in the distribution of beliefs between self-identified political orientations indicates that traditional political ideologies have some bearing on how participants relate to blockchain technology. For example, the "left-of-center" group articulated distinct beliefs about economic opportunity, fairness of wealth distribution, privacy, and the growth of crypto (Q11-13, Q6, and Q9), suggesting that left-of-center respondents are more likely to apply more community-oriented multi-stakeholder values to the blockchain ecosystem. The "nonaligned" group, on the other hand, articulated distinct beliefs about gender equity and government regulation (Q15 and Q5), suggesting this group is more clearly defined by lower trust in existing government institutions[2]. This lower approval of government regulation suggests that those who identified as neither left-of-center nor right-of-center are more likely to position themselves as separate from existing political and governance systems entirely.

We are interested in understanding the extent to which the characteristics of developers and users of specific blockchains are distinctive of each blockchain. Critics like David Golumbia have argued that Bitcoin, both in its design and ideological constitution, is principally a conservative movement interested strictly in Bitcoin's record of gaining value (Golumbia 2016). Those observations were not made in opposition to Ethereum or any other blockchain, although Ethereum had been live for a year at the time of Golumbia's writing. Our findings indicate that in fact, there are few differences between Bitcoin and Ethereum users. However, differences in technical implementation between Bitcoin and Ethereum may relate to differences in opinion on their governance. Unlike Bitcoin, which has limited support for transactions other than money transfers, Ethereum as an infrastructure enables the developing and building of various applications and projects. The broader set of use cases for Ethereum may lead its users to believe a broader set of stakeholders should be involved in its governance. Furthermore, Ethereum affiliation was associated with a greater sensitivity to perceived socioeconomic inequity, which may relate to differences in how the blockchains are used alongside other technology. In Ethereum ecosystems, users linking their own blockchain activity to other personally-identifying information, such as Discord handles or Twitter accounts, is not uncommon; more research is needed to understand whether lower rates of anonymity relate to greater awareness of actual or perceived social demographics.

---

[2] We refer in this work to respondents who chose to identify as neither left-of-center nor right-of-center as "nonaligned". We choose this term in contrast to a term such as "apolitical" in a nod to ideas of political agnosticism developed by ethnographers in observing open-source communities (Gabriella Coleman 2013).

Communities organized around crypto are proving to be a laboratory for new ways that humans can organize collective action, but are not operating in a historical vacuum: it appears that some patterns observed in early users of other internet technologies have arisen or continue to appear in the blockchain context as well. To complement this sociological work, further anthropological research could shed some light on the extent to which the economic and political beliefs held by participants in crypto echo the ideologies of two earlier movements: the cryptographic hacker and open-source software communities. The distribution of responses relating to fair rewards for developer teams and the utility of hard work and in crypto indicates that meritocratic values are prevalent; meritocracy may play a similar role in blockchain ecosystems, themselves often open source, as it has in prior open source and hacker communities (Gabriella Coleman 2013; Dunbar-Hester 2019). Privacy has been at the forefront of concerns in the development of internet technologies since the cypherpunks (Hughes 1993) and remains prevalent in blockchain (Brunton 2020). Further research is needed to understand how these values compare to those of open-source software communities and early adopters of the internet or how they may have changed over time as cryptocurrencies become more mainstream.

## 6     Limitations

### 6.1    Survey methodology

Selection bias may arise given that the random sample assumption is limited by how the survey was distributed. In particular, since the survey was opt-in, people with stronger and potentially more extreme opinions may have been more motivated to complete the survey. Also, the survey was made available only in English, and so is likely not representative of the full geographic distribution of users of blockchain technologies. Presentation of some preliminary findings prior to finalizing data collection may also have influenced some respondents. Additionally, we do not have a guarantee of uniqueness of each respondent; moreover, the two recruitment strategies we used may have motivated respondents to provide multiple responses.

In choosing the wording of each question and answer choice, we made an effort to mitigate response bias. Still, we have identified some limitations in interpreting questions based on the wording of the questions. Q5 may have had an insufficient distinction between the two most commonly-selected choices. In answering Q14, respondents who selected "Keep on doing what we're doing" may have rejected the premise of the question rather than believed this was a way to achieve the stated goal. Additionally, while we intended Q15 to refer to perceptions of gender inequity in participation or compensation within crypto, the wording of the choices may have been too vague.

Further demographic information would be valuable context for interpreting some questions. Future work could include a question on the geographic location of respondents, where local regulations and political attitudes could inform a more detailed analysis of questions on national government regulation and political affiliation. Interpretation of question 18 on political self-identification may be similarly limited by differences in how terms such as "liberal" and "conservative" are understood across the world.

## 6.2 Analysis

To be able to use PCA for the discrete data, we one-hot encoded specific choices. While PCA is generally better suited to continuous data than boolean data, we find that in this context the results were cleanly interpretable. We also chose not to include null responses as an additional coded choice for feature selection or factor analysis. While this does result in using only a subset of the responses and potentially removing relevant information about respondents' beliefs, it prevents the null responses from receiving artificially high importance due to their relative rarity and bypasses the difficulty in interpreting the null response.

## 7 Conclusion

In this work, we have introduced a new survey of blockchain users' political, economic, and governance opinions with respect to crypto. Based on 3710 survey responses, we find that users were spread across a variety of perceptions of the distribution of economic power, normative beliefs about the distribution of power in governance, and opinions on the role of external regulation of crypto, though they were broadly in agreement that crypto has a net-positive impact on the world. Equal numbers of respondents self-identified as liberal or non-aligned, while only about a third as many respondents self-identified as conservative; this self-reported political affiliation was associated with differences in opinions on most questions, but especially on economic fairness, decision-making power, and how to obtain favorable regulation. In contrast, we observed few differences in opinions between respondents affiliated with Bitcoin and with Ethereum, on issues of blockchain governance and regulation and on personal attitudes towards crypto. While the full field of beliefs elides neat interpretation in terms of underlying factors, we found that the existence of a political dimension was supported both by a theory-driven construct and by a common, well-validated analytical method (PCA).

While this dataset is an important step towards understanding the distribution of crypto users' beliefs about blockchain technology and its utility, open questions remain as to why users believe what they do about crypto and how their beliefs match up with reality. For example, considering the question of who should have decision-making authority over a blockchain: Is the large-minority opinion that token-holding voters should control a blockchain underlied by a belief that minimizing human input to governance will make it more efficient and less flawed? Is there a disconnect between the common normative beliefs of what should be happening in cryptogovernance and which types of stakeholders actually can and do participate in governance of major blockchains?

Although our research found only a few instances where affiliation with a specific blockchain was associated with differences in beliefs, further research is needed to better understand whether specific architectures or ecosystems within crypto differ in the values or goals embedded in them. Future interdisciplinary work could shed some light on the extent to which participants have common understandings of core signifiers such as decentralization and autonomy.

Given that our work takes inspiration from the long-running Pew political survey, we see the need for a regular survey of cryptopolitical sentiment, with an added demographic panel. This could facilitate the identification and comparison of ideologies and modes of participation within newer chains such as Solana, L2s such as Polygon, and even large DAOs.

## 8  Conflict of Interest

The authors declare that the research was conducted in the absence of any commercial or financial relationships that could be construed as a potential conflict of interest.

## 9  Author Contributions

LMK conducted data analysis and wrote manuscript. SF advised on methodology and advised on manuscript. JT designed and deployed the survey instrument and advised on manuscript.

## 10  Funding

Lucia Korpas and Joshua Tan were supported by grants from the Filecoin Foundation and from One Project.

## 11  Acknowledgments

The authors would like to acknowledge Michael Zargham for technical discussion, Ann Brody for qualitative discussion, and Tyler Sullberg and Nathan Schneider for feedback on the manuscript.

## 12  Data Availability Statement

The dataset generated and analyzed for this study can be found in the Metagovernance Project's Govbase repository on Airtable (https://airtable.com/shrgnUrj0dqzZDsOd/tblvwbt4KFm8MOSUQ/viw82nVNrdHFrowoo). The Python code used to conduct the analysis and produce the figures can be found in the GitHub repository for this work (https://github.com/metagov/cryptopolitics-paper).

# 13 References

Bauerle, Nolan, and Peter Ryan. 2018. "CoinDesk Releases Q2 2018 State of Blockchain Report." CoinDesk. July 25, 2018. https://www.coindesk.com/business/2018/07/25/coindesk-releases-q2-2018-state-of-blockchain-report/.

Beltramini, Enrico. 2021. "The Cryptoanarchist Character of Bitcoin's Digital Governance." *Anarchist Studies* 29 (2): 75–99.

Brody, Ann, and Stéphane Couture. 2021. "Ideologies and Imaginaries in Blockchain Communities: The Case of Ethereum." *Canadian Journal of Communications* 46 (3): 19 pp-19 pp.

Brunton, Finn. 2020. *Digital Cash: The Unknown History of the Anarchists, Utopians, and Technologists Who Created Cryptocurrency*. Princeton University Press.

Buchholz, Katharina. 2021. "These Are the Countries Where Cryptocurrency Use Is Most Common." World Economic Forum. February 18, 2021. https://www.weforum.org/agenda/2021/02/how-common-is-cryptocurrency/.

Chohan, Usman W. 2017. "Cryptoanarchism and Cryptocurrencies." https://papers.ssrn.com/sol3/papers.cfm?abstract_id=3079241.

De Filippi, Primavera, and Benjamin Loveluck. 2016. "The Invisible Politics of Bitcoin: Governance Crisis of a Decentralized Infrastructure." *Internet Policy Review* 5 (4). https://papers.ssrn.com/sol3/papers.cfm?abstract_id=2852691.

Dunbar-Hester, Christina. 2019. *Hacking Diversity*. Princeton University Press. https://press.princeton.edu/books/hardcover/9780691182070/hacking-diversity.

Gabriella Coleman, E. 2013. *Coding Freedom: The Ethics and Aesthetics of Hacking*. Princeton University Press.

"Global State of Crypto, 2022." 2022. *Gemini*. Gemini. https://www.gemini.com/state-of-us-crypto.

Golumbia, David. 2016. *The Politics of Bitcoin: Software as Right-Wing Extremism*. U of Minnesota Press.

Hughes, Eric. 1993. "A Cypherpunk's Manifesto." March 9, 1993. https://www.activism.net/cypherpunk/manifesto.html.

Husain, Syed Omer, Alex Franklin, and Dirk Roep. 2020. "The Political Imaginaries of Blockchain Projects: Discerning the Expressions of an Emerging Ecosystem." *Sustainability Science* 15 (2): 379–94.

Liu, Yue, Qinghua Lu, Liming Zhu, Hye-Young Paik, and Mark Staples. 2021. "A Systematic Literature Review on Blockchain Governance," May. https://doi.org/10.48550/arXiv.2105.05460.

May, Timothy C. 1994. "THE CYPHERNOMICON: Cypherpunks FAQ and More." September 1994. https://nakamotoinstitute.org/static/docs/cyphernomicon.txt.

Nadeem, Reem. 2021. "Beyond Red vs. Blue: The Political Typology." Pew Research Center. https://www.pewresearch.org/politics/2021/11/09/beyond-red-vs-blue-the-political-typology-2/.

Nakamoto, Satoshi. 2008. "Bitcoin: A Peer-to-Peer Electronic Cash System." https://bitcoin.org/bitcoin.pdf.

Pelt, Rowan van, Slinger Jansen, Djuri Baars, and Sietse Overbeek. 2021. "Defining Blockchain Governance: A Framework for Analysis and Comparison." *Information Systems Management* 38 (1): 21–41. https://doi.org/10.1080/10580530.2020.1720046.

Perrin, Andrew. 2021. "16% of Americans Say They Have Ever Invested in, Traded or Used Cryptocurrency." Pew Research Center. https://www.pewresearch.org/fact-tank/2021/11/11/16-of-americans-say-they-have-ever-invested-in-traded-or-used-cryptocurrency/.

Reijers, Wessel, Fiachra O'Brolcháin, and Paul Haynes. 2016. "Governance in Blockchain Technologies & Social Contract Theories." *Ledger* 1 (December): 134–51. https://doi.org/10.5195/ledger.2016.62.

Ryan, Peter. 2018. "Left, Right and Center: Crypto Isn't Just for Libertarians Anymore." *CoinDesk*, July 2018. https://www.coindesk.com/markets/2018/07/27/left-right-and-center-crypto-isnt-just-for-libertarians-anymore/.

Szabo, Nick. 1996. "Smart Contracts: Building Blocks for Digital Markets." *EXTROPY: The Journal of Transhumanist Thought* 18 (2).

Winner, Langdon. 1980. "Do Artifacts Have Politics?" *Daedalus*, Winter 1980.

Zamfir, Vlad. 2019. "Against Szabo's Law, For A New Crypto Legal System." *Crypto Law Review* (blog). January 2019. https://medium.com/cryptolawreview/against-szabos-law-for-a-new-crypto-legal-system-d00d0f3d3827.

# 14     Tables

## Table 1

| Question number | Question text | Choice text | Contribution to political score |
|---|---|---|---|
| 1 | Which statement comes closest to your views? | There is one (layer 1) blockchain that is the best. | |
| | | There is no one best blockchain. | |
| 2 | Which blockchain is the best? | Bitcoin | |
| | | Ethereum | |
| | | Solana | |
| | | Cardano | |
| | | Polkadot | |
| | | Other | |
| 3 | Which statement comes closest to your views? | Crypto is mainly an economic technology. | |
| | | Crypto is mainly a political philosophy and/or lifestyle. | |
| 4 | Which statement comes closest to your views? | My goal in crypto is to have fun. | |
| | | My goal in crypto is to make as much money as possible. | |
| | | My goal in crypto is to create social change and/or disrupt the industry. | |
| | | My goal in crypto is to earn a living and/or build my career. | |
| 5 | Which statement comes closest to your views? | Most or all cryptogovernance should be on-chain. | |
| | | Most or all cryptogovernance should be off-chain. | |
| | | Crypto does not need (human) governance; let the algorithms run as they were designed. | |
| | | However crypto governs itself, it should also be regulated by the government. | |
| 6 | Which statement comes closest to your views? | Privacy is the most important feature of blockchain and crypto. | 2 |
| | | Privacy is nice, but it's not the most important feature of blockchain and crypto. | 0 |
| 7 | Which statement comes closest to your views? | Government regulation of crypto will almost always do more harm than good. | 1 |
| | | Government regulation of crypto can do some good, e.g. it can help force blockchains to become more decentralized. | |
| | | Government regulation of crypto is critical to protect the public interest in these technologies. | -1 |
| 8 | Which statement comes closest to your views? | Having a central bank run a cryptocurrency is a good idea. | |
| | | Having a central bank run a cryptocurrency is a bad idea. | |
| 9 | In order to grow, the crypto ecosystem should: | Build art and community. | -1 |
| | | Help people around the world earn a living. | -1 |
| | | Build useful tech that solve real problems for a set of users. | 1 |
| | | Provide financial instruments for maximum wealth creation. | 1 |
| 10 | Which statement comes closest to your views? | Blockchain and DeFi are beneficial technologies that, on balance, will help most members of society. | |
| | | Blockchain and DeFi are predatory technologies that, on balance, will harm most members of society. | |
| 11 | Which statement comes closest to your views? | Most crypto teams make a fair and reasonable amount of profit. | 1 |
| | | Crypto teams make too much profit. | -1 |
| 12 | Which statement comes closest to your views? | The economic system in crypto is generally fair to most of its participants. | 1 |
| | | The economic system in crypto unfairly favors powerful interests. | -1 |
| 13 | Which statement comes closest to your views? | Most people who want to get ahead in crypto can make it if they're willing to work hard. | 1 |
| | | In crypto, hard work and determination are no guarantee of success for most people. | -1 |
| 14 | To get more favorable regulation of cryptocurrencies from national governments, the most important thing the crypto community can do is: | Adapt our technology and practices in order to minimize potential conflicts with the law. | |
| | | Work hand-in-hand with regulators to identify a solution that works for both government and industry. | |
| | | Hire lawyers and lobbyists; organize the community to mount a public pressure campaign on politicians. | |
| | | Keep on doing what we're doing, legal or not. | 1 |
| 15 | Which statement comes closest to your views? | Crypto has a gender problem. | 0 |
| | | Crypto does not have a gender problem. | 1 |
| 16 | Who should have decision-making power over a blockchain? | The public, elected representatives, and/or national leaders | |
| | | A wide variety of on- and off-chain stakeholders including token holders, node operators, application developers, foundations, and users | |
| | | The token holders and/or node operators, i.e. voters, as determined by the protocol | |
| | | The core developers and technical staff of a blockchain | |
| 17 | I'm here for... | the memes | |
| | | the jobs | |
| | | the tech | |
| | | the airdrops | |
| 18 | I identify as: | Liberal or left-wing | -1 |
| | | Conservative or right-wing | 1 |
| | | Neither | 0 |

| | 19 | OPTIONAL: Do you affiliate with any of the following ecosystems or communities? | Bitcoin | |
| | | | Ethereum | |
| | | | Solana | |
| | | | Cardano | |
| | | | Polkadot | |
| | | | Other | |

**Table 2**

| Politics score | Assigned type |
| --- | --- |
| [-7,9] | Overall possible range of scores |
| >= 5 | Crypto-anarchocapitalist |
| 0 < x < 5 | Crypto-libertarian |
| x = 0 | Crypto-centrist |
| -3 < x < 0 | Crypto-communitarian |
| <= -3 | Crypto-leftist |

# Supplementary Material

## 1 Supplementary Information

### 1.1 Recruitment strategy: Ad-hoc typology

Based on their responses to the survey, each respondent was assigned a political faction, an economic faction, and a governance class. These were computed immediately upon completion of the survey according to numerical weights assigned to each answer choice and formulas defined for each faction. The faction definitions and names were developed with input from the community, and served largely as a way to (1) recruit participants and (2) help participants interpret their results.

Depending on the response to Question 3 – whether the respondent considered crypto to be primarily a political philosophy or an economic technology – each respondent was correspondingly presented with either their political faction or economic faction as their overall assigned faction. The five political factions ("crypto-leftist", "DAOist", "true neutral", "crypto-libertarian", and "crypto-ancap") were an alternate naming scheme for our constructed political axis. The five economic factions ("earner", "cryptopunk", "NPC", "techtrepreneur", "degen") were intended to represent the respondents' primary mode of economic engagement with crypto. Additionally, we defined four "classes" (Szabian, Gavinist, Zamfirist, and Walchian) that were intended to capture respondents' beliefs about governance and government regulation, inspired by the positions articulated by Nick Szabo, Gavin Wood, Vlad Zamfir, and Angela Walch.

The mechanism for assigning the factions based on respondents' answers was a weighted sum of their responses to each question; every answer choice in every question adds points to one or more factions, and the faction assigned is the one with the most points, with some thresholding to account for respondents who did respond to all questions. In cases where not enough responses were given, the assignment defaulted to "true neutral" or "NPC". Questions 6, 7, 9, 11, 12, 13, 14, 15, and 18 were used in computing the political faction, questions 4, 9, 17 in computing the economic faction, and questions 5, 7, 8, 14, and 16 in computing the governance class.

The political faction names and definitions were conceptualized and iterated through a community-based effort at crypto conferences and online, especially within the Metagovernance Project Slack community. The names of the factions reflected both popular slang in crypto (degen, cryptopunk, crypto-libertarian, crypto-leftist, DAOist, crypto-ancap) as well as a few inventions of the authors when there was no existing slang or word for that archetype (earner, techtrepreneur, true neutral, NPC). The visual representations of the factions, which were used as the underlying images of a series of NFTs, played off various memes (e.g. shiba inu / doge representing degens) and cultural icons (e.g. Shrek representing crypto-libertarians) common in crypto and on the internet more broadly.

Within the main text, the political "factions" were renamed to the described "types".

### 1.2 Differences in assigned political types by self-reported political orientation and blockchain ecosystem affiliation

There was a statistically significant difference in the overall distribution of assigned cryptopolitical factions across the left-of-center, right-of-center, and nonaligned groups. Right-of-center respondents were the only group more likely to be assigned the crypto-anarcho-capitalist faction than the crypto-libertarian faction, and less likely than the other two groups to be assigned true neutral,

DAOist, or crypto-leftist. Unlike the other two groups, left-of-center respondents were more likely to be assigned any of crypto-centrist, crypto-communitarian, or crypto-leftist than to be assigned crypto-anarcho-capitalist. Nonaligned respondents generally split the difference between the proportions of left-of-center and right-of-center respondents assigned each faction except for crypto-libertarian, which they were more likely to be assigned than either of the other two groups. Note that while Q18 was used directly in assigning the political faction, it was one of nine such questions, each with similar weights in the overall cryptopolitical score.

A higher percentage of Bitcoin-affiliated respondents than Ethereum-affiliated respondents received the "crypto-anarcho-capitalist" label.

## 1.3  Are there clusters of respondents or features?

We were interested in understanding whether the range of responses would be better captured by the idea of "types" (groupings of respondents, consistent with data-driven clustering methods) or "axes" (grouping of questions, consistent with factor analysis techniques). We investigated this by extending our PCA analysis with feature agglomeration, based on a feature set of 48 choices provided across 17 questions.

We used feature agglomeration to hierarchically generate clusters of features, where each feature corresponded to the selection of a provided choice. Given that these features are boolean in nature, we used the Dice distance metric to determine distances between features; this can be understood as the fraction of the sample for which the two features intersect or overlap, as a representation of the shared information between them. To determine which clusters should be merged and when, we used the complete linkage criterion, which relates to the maximum distances between all features of the two clusters.

From the feature agglomeration results, we ask whether the assigned political types directly correlate with any of the identified factors (i.e., to what extent does our categorization of beliefs as political, economic, or governance-related meaningfully describe joint variations in the choices selected?). At the threshold where three clusters were present, the clusters contained A=31, B=10, and C=7 choices. The cluster that shared the most features in common with the important principal component features listed above was the cluster with 10 features: the specific choices listed above for beliefs on economic fairness, gender, and political affiliation (Q11-Q13, Q15, and Q18) appeared in this feature. The two specific choices relating to government regulation (Q5 and Q7) were present together in the cluster with 7 features Although cluster B aligned fairly well with the first (politically-focused) principal component, our exploration of the other clusters, and the hierarchy that produced them, did not indicate much qualitative support for this approach. This null assessment is supported by our mapping of respondents into feature space: both factor analysis methods shown in [Supplementary Figure S7](#) shows responses organized around a single centroid, not the multiple clusters that would be expected in the presence of clear respondent types.

Given that the feature agglomeration method allowed us to traverse the entire hierarchy of feature clusters, we also looked at the agglomerated clusters to see whether three stable feature clusters arise from the full feature set. Based on the distances required for clusters to merge, it appears instead that five distinct latent variables may best describe the distribution of respondents' beliefs, as shown in [Supplementary Figure S8](#). That said, the smaller two out of the three clusters identified above persisted unchanged at the five-cluster threshold, suggesting that the underlying factors for each of those two clusters have more explanatory power.

# 2 Supplementary Figures and Tables

## 2.1 Supplementary Figures

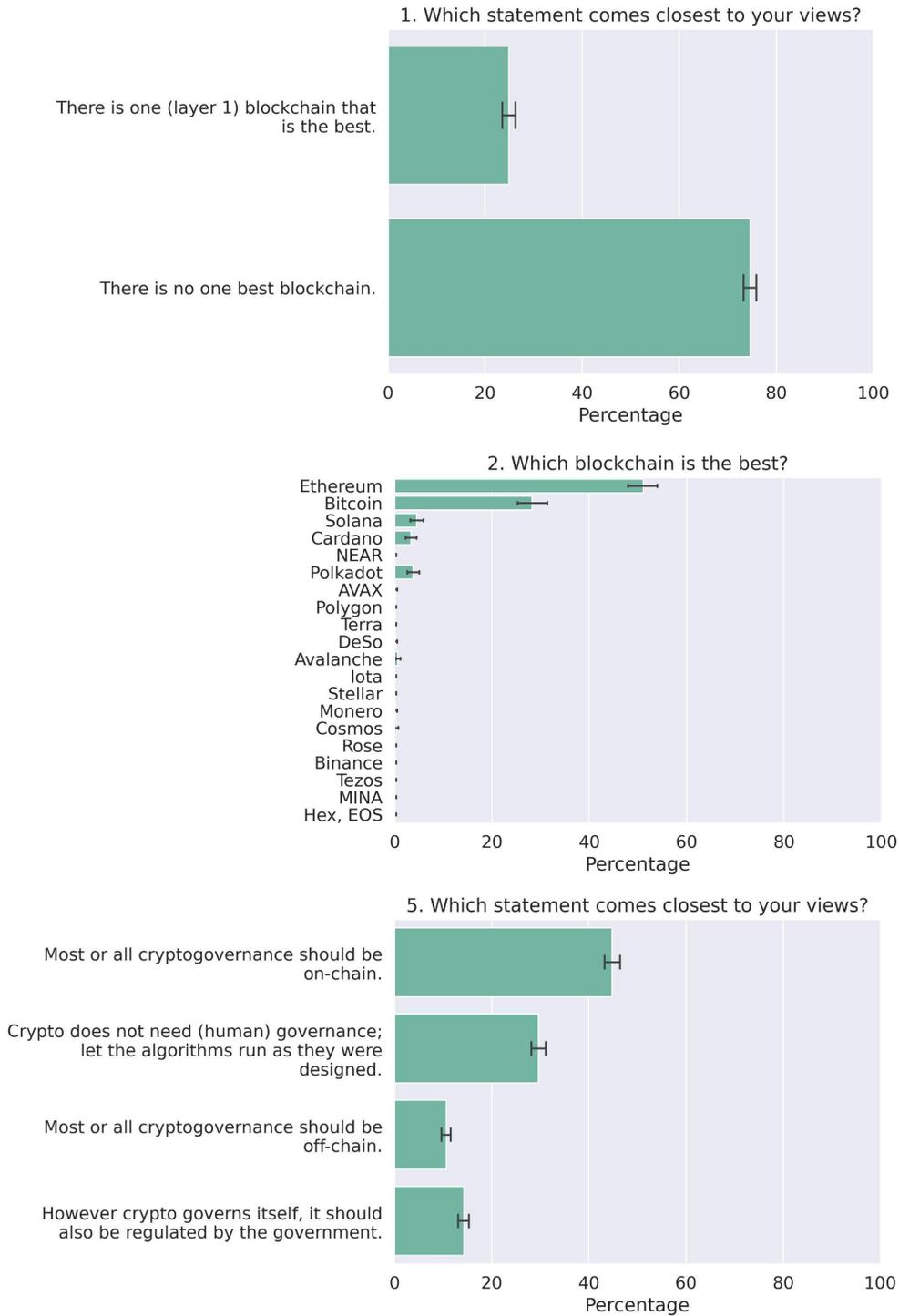

**Supplementary Figure S1**. Distribution of responses to questions 1, 2, and 5. Note that question 2 was presented only to respondents who selected the response "There is one (layer one) blockchain that is the best" for question 1, and that respondents were given the option to select Ethereum, Bitcoin, Solana, Polkadot, Cardano, or a fill-in-the-blank "Other".

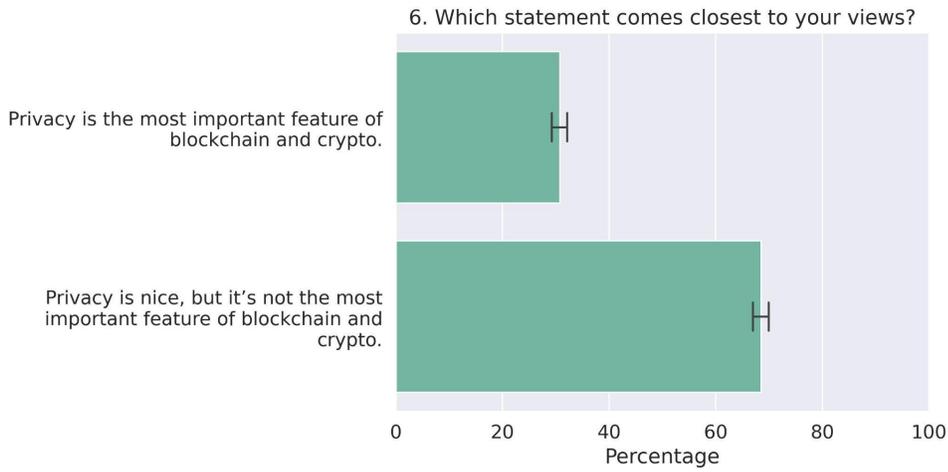

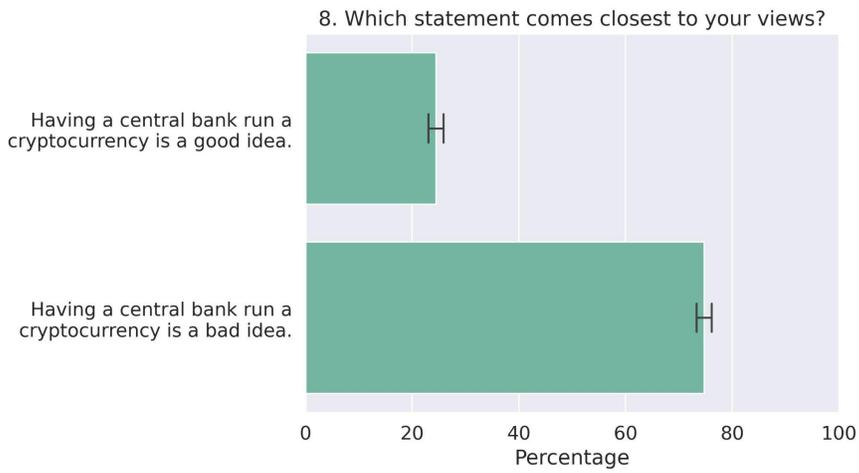

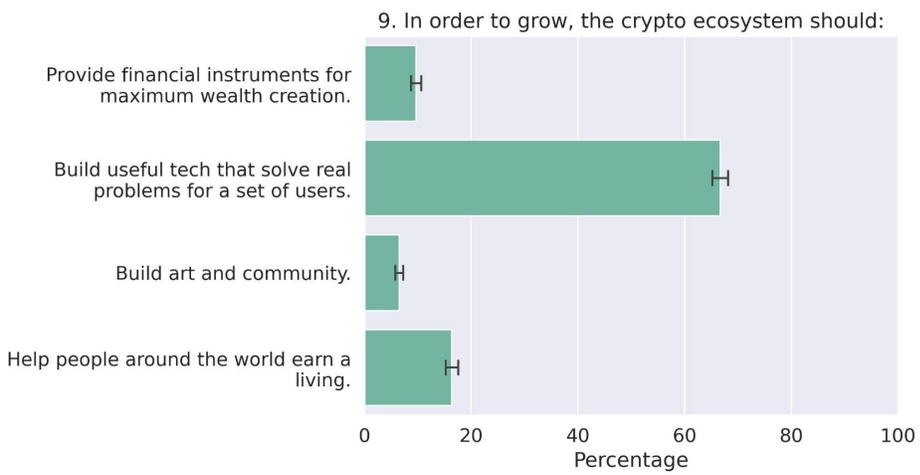

**Supplementary Figure S2**. Distribution of responses to questions 6, 8, and 9.

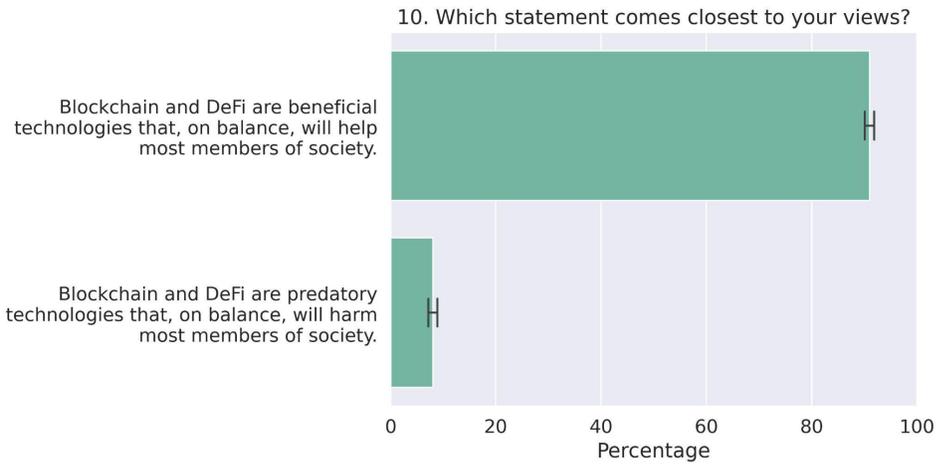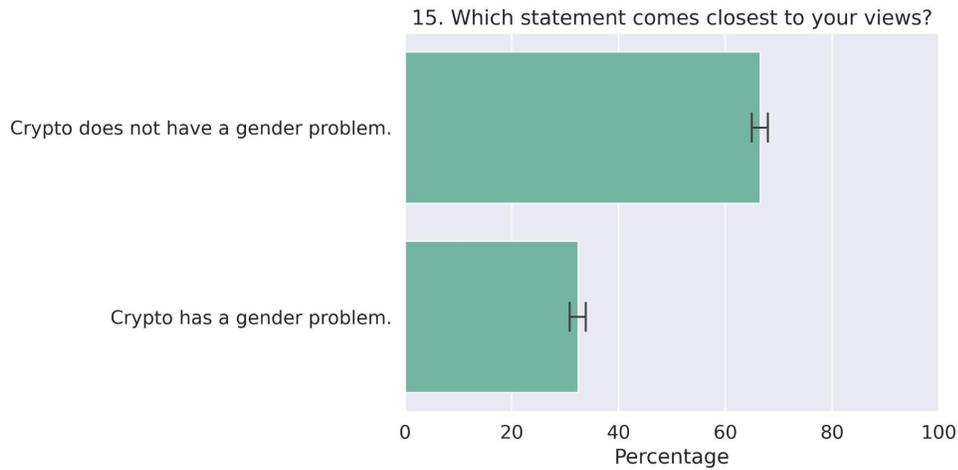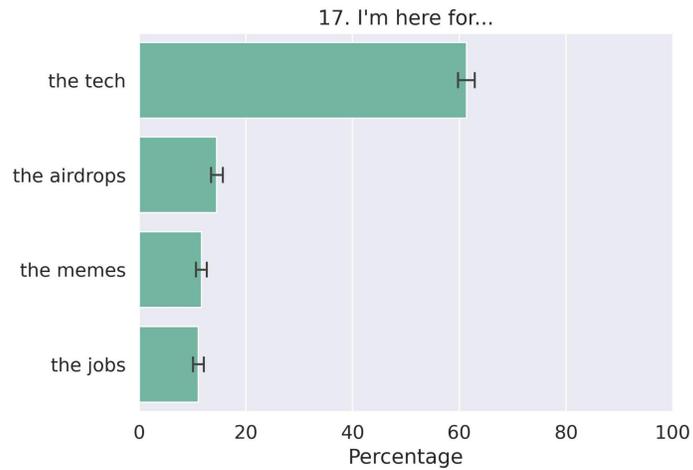

**Supplementary Figure S3**. Distribution of responses to questions 10, 15, and 17.

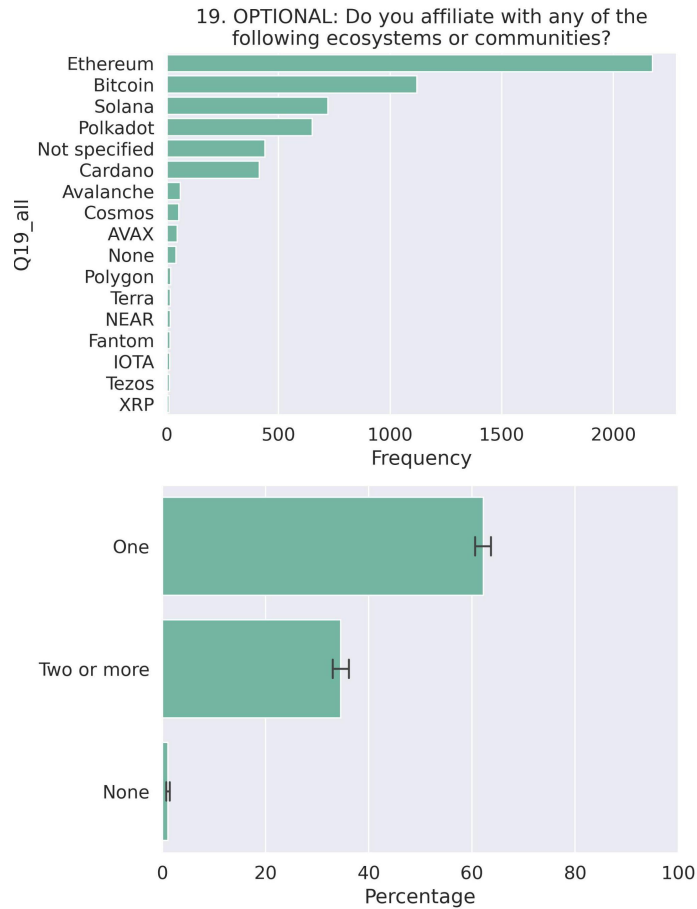

**Supplementary Figure S4**. Distribution of responses to question 19, and distribution in the number of those multiple-select choices that respondents who answered the question chose. Note that respondents were given the option to select Ethereum, Bitcoin, Solana, Polkadot, Cardano, or a fill-in-the-blank "Other". For plot legibility, only the ecosystems listed by at least 10 respondents are shown here.

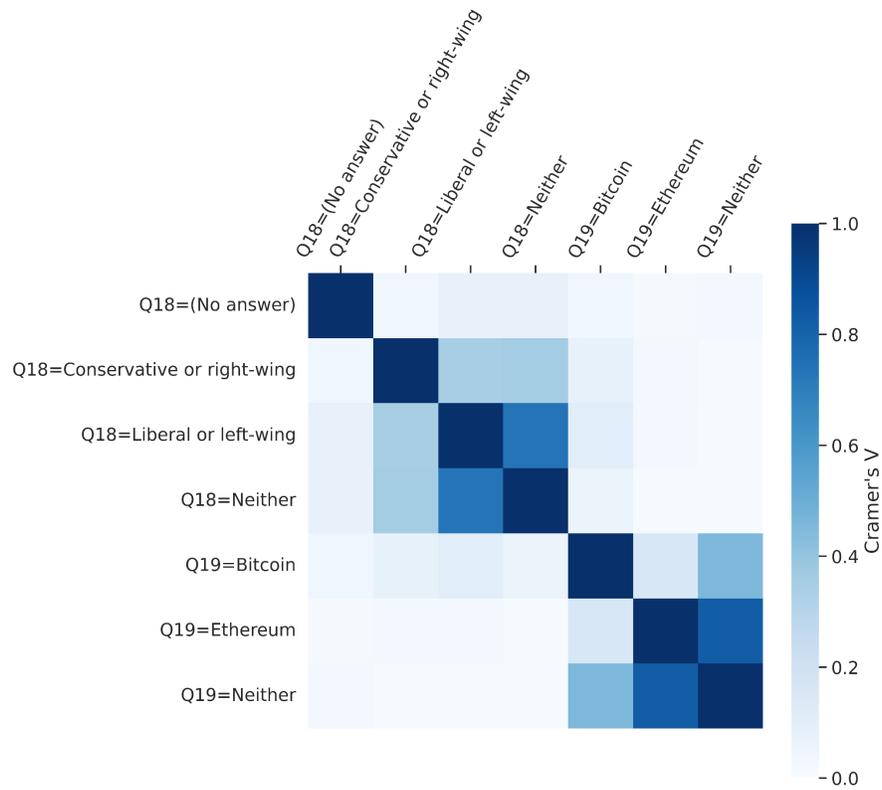

**Supplementary Figure S5.** Correlation (Cramer's V) between choices to questions 18 and 19. The low correlation between choices belonging to different questions indicates that political self-identification and blockchain affiliation are largely independent.

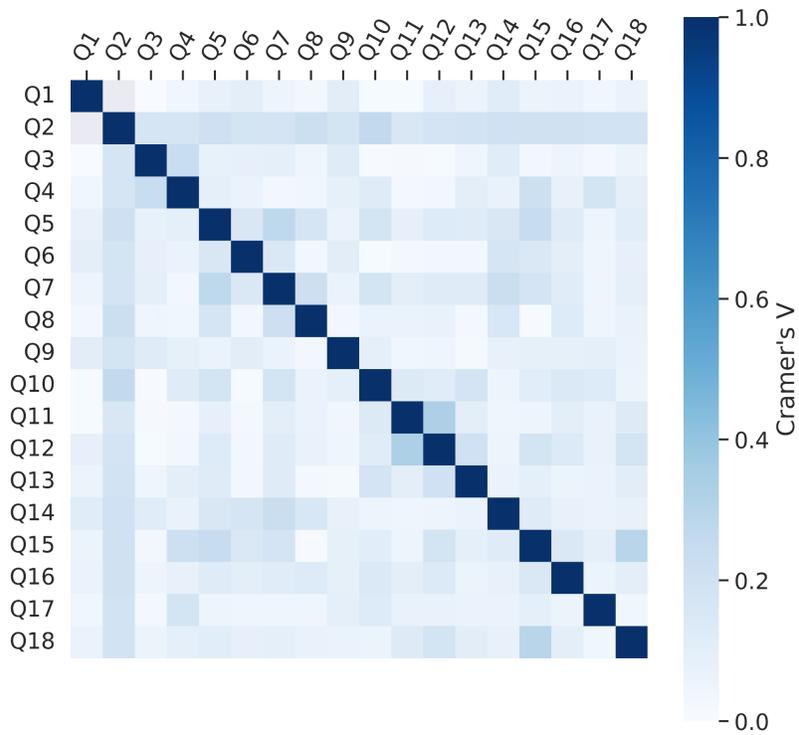

**Supplementary Figure S6**. Correlation (Cramer's V) between all questions. Gray indicates that the correlation between questions was not computed; this is used when one question was only presented to the respondents upon a particular selection for the other question. No clusters are clearly observable: the questions are not systematically related in macro groupings.

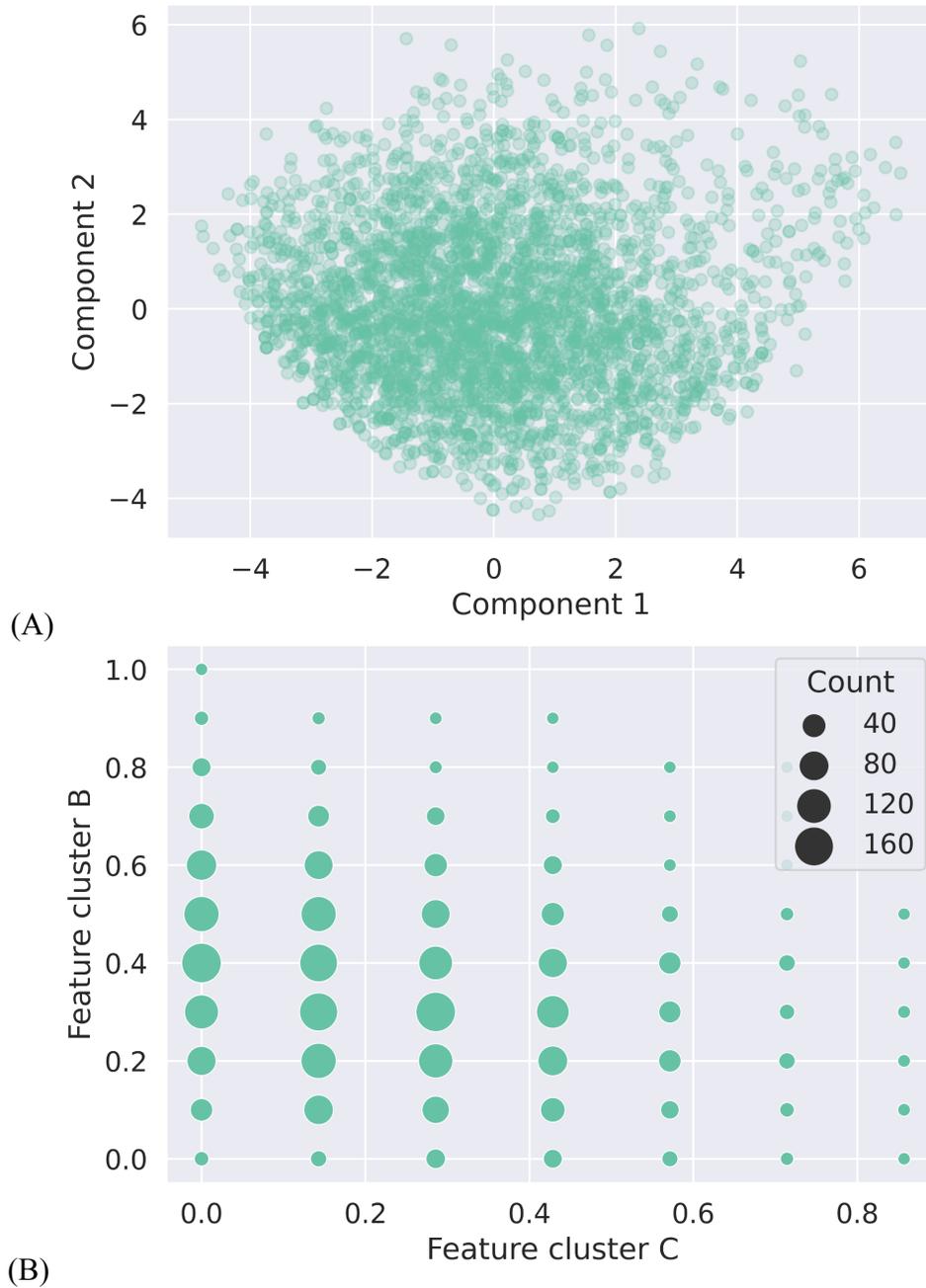

**Supplementary Figure S7.** (**A**) Projection of the respondents onto the first two principal components. (**B**) Projection of the respondents onto the feature clusters B (10 features) and C (7 features). In both cases, there is a continuous (rather than discrete) distribution of respondents along the two axes. The unimodal distribution of respondents along the two axes does not suggest the presence of clusters of respondents by their responses.

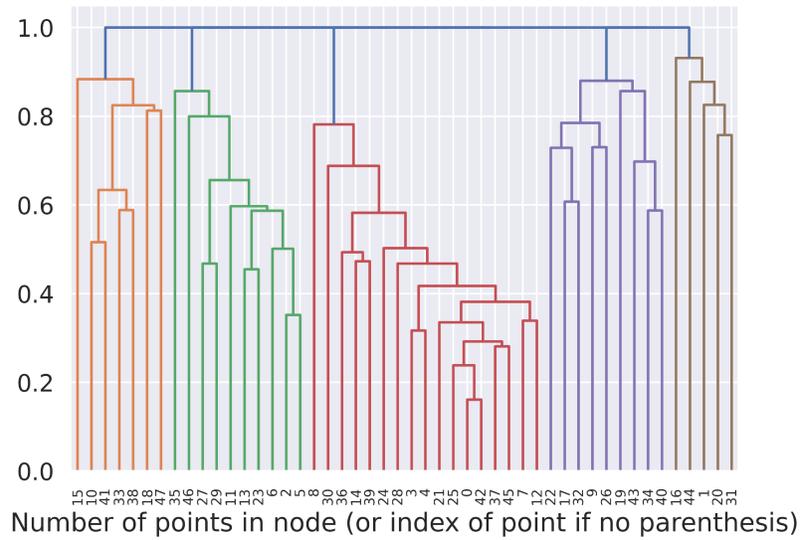

**Supplementary Figure S8.** Dendrogram describing the hierarchical agglomeration of feature clusters. The colors delineate the first five clusters that branch off from the one-cluster case (i.e., the one containing all features). The three-cluster case involves grouping some of these together.